\documentclass[aps, twocolumn, amsmath, amssymb, citeautoscript,longbibliography]{revtex4-1}
\usepackage{amsmath}
\usepackage{graphicx}
\usepackage{dcolumn}
\usepackage{natbib}

\begin{document}
\title{Quantum Stern-Gerlach experiment and path entanglement of a Bose-Einstein condensate}
\author{Mandip Singh}
\email{mandip@iisermohali.ac.in}
\date{\today}
\affiliation{Department of Physical Sciences
\\Indian Institute of Science Education and Research Mohali, Sector-81,S.A.S. Nagar, Mohali, 140306, India}

\begin{abstract}
In this paper, a quantum Stern-Gerlach thought experiment is introduced where, in addition to the intrinsic angular momentum of an atom, the magnetic field is considered to be a quantum mechanical field. A free falling spin polarised Bose-Einstein condensate passes close to a flux-qubit and interacts with the quantum superimposed magnetic field of the flux-qubit. Such an interaction results a macroscopic quantum entanglement of the path of a Bose-Einstein condensate with the magnetic flux quantum state of the flux-qubit. In this paper, three regimes of coupling between the flux-qubit and a free falling Bose-Einstein condensate are discussed. This paper also explains, how to produce a path entangled Bose-Einstein condensate where, the condensate can be located at physically distinct locations simultaneously. This paper highlights new insights about the foundations of the quantum Stern-Gerlach experiment.
\end{abstract}
\maketitle

\section{Introduction}
  The Stern-Gerlach (SG) experiment has a fundamental significance as it provides a clear evidence of quantization of intrinsic angular momentum of an atom (historically known as space-quantization) \cite{sg1,sg2,sg3}. The classic SG experiment is regarded as one of the most important and extensively explored experiments of physics \cite{sgphys, feynman, scully, gilbert, gv, patil, prop, elec, dyna, zur}. In the context of foundations of quantum physics the SG experiment is widely regarded as a precursor of thought experiments. According to quantization of intrinsic angular momentum, the component of the angular momentum of a particle along an arbitrary fixed axis (also known as the quantization axis) is quantised. In a typical SG apparatus the collimated atoms having a nonzero magnetic moment are passed through the magnetic field gradient. The path of each neutral atom is deflected by a spin projection dependent force. If the initial quantum state of an atom is a quantum superposition of spin projections then after passing through the SG apparatus the path of the atom is quantum entangled with the spin. For many such noninteracting atoms the quantum entanglement persists at a single atom level only. In the SG experiment, the magnetic field is regarded as a classical field and the spin degree of freedom is quantised.

 In this paper, a quantum SG thought experiment is presented where, in addition to the spin of an atom, the magnetic field obeys the laws of quantum mechanics. In the quantum SG experiment, the magnetic field can exist in a quantum superposition state which leads to remarkable consequences for example, if a spin polarized Bose-Einstein condensate (BEC) \cite{bec2,bec1,chip,revbec} is passed through a quantum SG apparatus then the path of the BEC can be quantum entangled with the magnetic field. Such an entangled quantum state is a multi-particle macroscopic entangled quantum state or a Schr\"{o}dinger-cat state \cite{schd}. In this paper, a proof-of-principle idea of the quantum SG thought experiment is presented and its experimental feasibility is  discussed. This paper also highlights fundamental conditions to realise the quantum SG experiment and a path entangled BEC.

\section{Stern-Gerlach experiment and quantum entanglement}
Consider a neutral atom of total spin $\mathbf{F}$ is passed through a semi-classical SG apparatus, which has a predominant component of the magnetic field and a predominant component of its gradient along the $z$-axis. The $z$-axis is also considered to be the quantization axis.  The atom is moving along the $x$-axis with a zero expectation value of its transverse linear momentum (along the $z$-direction) prior to its interaction with the magnetic field gradient. The projection of the angular momentum along the quantization axis is $m_{F}\hbar$, where $m_{F}$ is an integer that varies from $-F$ to $+F$ and $\hbar$$(h/2\pi)$ is the reduced Plank 's constant. Therefore, the total quantum state of the moving atom at time $t$ in the centre-of-mass frame is $|\psi(\mathbf{r};t)\rangle \sum_{m_{F}=-F}^{F} c_{m_{F}}(t) |m_{F}\rangle$, where $|\psi(\mathbf{r};t)\rangle$ is the initial quantum state, in the external degrees of freedom, of the atom in the centre-of-mass frame prior to its interaction with the magnetic field. In this paper, the bold symbols placed in the argument of ket vectors denote the basses in which a ket vector is expressed except the time parameter $t$, which is separated by a semicolon. Therefore, $|\psi(\mathbf{r};t)\rangle=\int \psi(r,t)|r\rangle \mathrm{d}r$, where $|r\rangle$ is the position basis and  $\psi(r,t)=\langle r|\psi(\mathbf{r};t)\rangle$ is the spatial wave function.
The quantum state $\sum_{m_{F}=-F}^{F} c_{m_{F}}(t) |m_{F}\rangle$ represents the spin degrees of freedom of an atom,  $|m_{F}\rangle$ is the quantum state corresponding to the projection of spin along the quantization axis with probability amplitude $c_{m_{F}}(t)$. As soon as the atom enters the magnetic field gradient region, a quantized force acts on the atom, which imparts to the atom a spin projection dependent momentum along the $z$-axis. The interaction term of the Hamiltonian of an atom of magnetic moment $\boldsymbol\mu$ in the presence of magnetic field $\mathbf{B}(r)$ is $-\mathbf{\boldsymbol\mu}.\mathbf{B}(r)$. Therefore, the force acting on the atom is $\mathbf{f}=\nabla(\boldsymbol\mu.\mathbf{B}(r))$, which can also be expressed as $\mathbf{f}=-g_{F}\mu_{B}\nabla(\mathbf{F}.\mathbf{B}(r))/\hbar$, where $g_{F}$ is the Landg$\acute{e}$ \textit{g} factor and $\mu_{B}$ is the Bohr magneton. The magnetic moment and the total spin are related through $\mathbf{\boldsymbol\mu}=-g_{F}\mu_{B}\mathbf{F}/\hbar$.
 The atom interacts with the magnetic field for a time $\Delta t$ during its passage through the SG apparatus. Therefore, the transverse linear momentum imparted to the atom, along the $z$-axis, is $p_{z}= - m_{F}g_{F}\mu_{B}\frac{\partial B_{z}}{\partial z} \triangle t$. The transverse linear momentum imparted to the atom is quantized because it is directly proportional to the quantised projection of the spin angular momentum along the quantization axis. The transverse linear momentum splitting increases with the strength of the magnetic field gradient. Consider an atom in its initial state $|\psi(\mathbf{r};t)\rangle\sum_{m_{F}=-F}^{F} c_{m_{F}}(t) |m_{F}\rangle$. After the interaction time $\triangle t$, total quantum state of the atom can be written as
\begin{multline}
\label{ent1}
  |\alpha(\mathbf{r}, \mathbf{m_{F}};t)\rangle_{a}=\sum_{m_{F}=-F}^{F} \left(\int \Psi(r, m_{F},t)|r\rangle \mathrm{d}r \right)\times\\
  c_{mF}(t) |m_{F}\rangle
\end{multline}
  The probability amplitude $\Psi(r, m_{F},t)$ is the spatial wave function of the atom with an imparted transverse linear momentum $p_{z}$ corresponding to $m_{F}$. Denote the term given in the bracket of Eq.~\ref{ent1} \emph{i.e.} $\int \Psi(r, m_{F},t)|r\rangle \mathrm{d}r$ with $|\psi_{p_{z}}(\mathbf{r};t)\rangle$, which corresponds to a quantum state of an atom with an imparted transverse linear momentum $p_{z}$.
   If the magnetic field gradient strength is sufficiently high to splits the atomic paths then $\langle\psi_{p'_{z}}(\mathbf{r};t)|\psi_{p_{z}}(\mathbf{r};t)\rangle = 0$, where $p_{z}$ corresponds to a given $m_{F}$ and $p'_{z}$ corresponds to $m'_{F}$ such that $m'_{F}$ is an allowed integer nearest to $m_{F}$. In this case the difference of the imparted transverse linear momenta corresponding to successive $m_{F}$ is greater than the uncertainty of the transverse linear momentum component of the atom 's initial wave function $\psi(r,t)$.
  Therefore, the quantum state given in Eq.~\ref{ent1} is a single atom entangled quantum state of spin projections and transverse linear momentum \emph{i.e.} the wave function $\Psi(r, m_{F},t)$ cannot be written as a product of a wave function of $m_{F}$ and a wave function of space variables.
 Because a quantised force acts on each atom independent of the quantum state of the other atoms  therefore, if more than one noninteracting atoms are passed through a semi-classical SG apparatus, the total quantum state will be a product state of a single atom entangled quantum state given in Eq.~\ref{ent1}. In the semi-classical SG experiment, the magnetic field is a classical field with well defined values and it is the intrinsic angular momentum of the atom which is quantised.

\section{Quantum Stern-Gerlach experiment: Principle}
\begin{center}
\begin{figure}
\begin{center}
\includegraphics[scale=0.20]{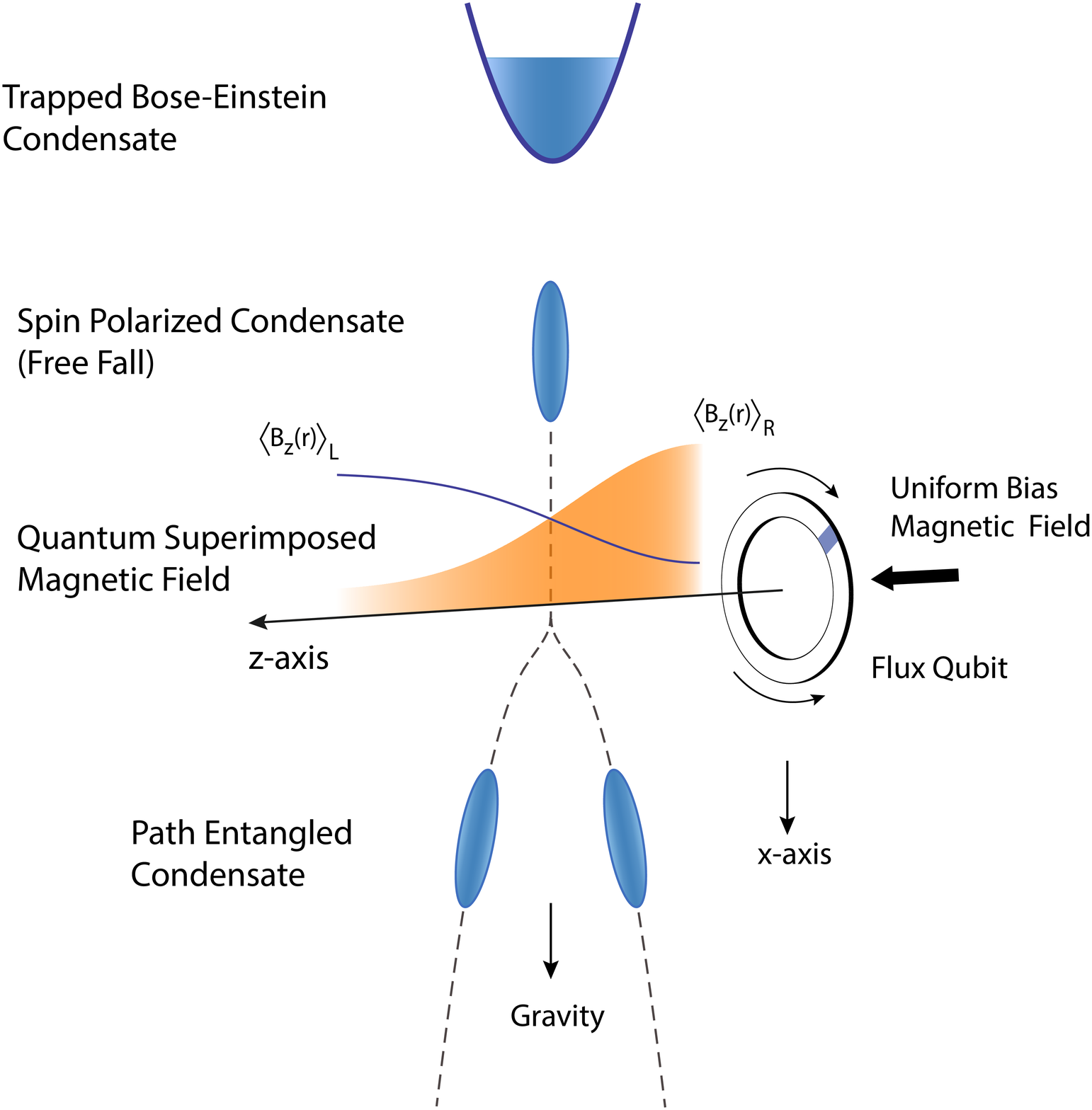}
\caption{\label{fig1} \emph{A schematic of the quantum Stern-Gerlach (SG) experiment, where a free falling spin polarised Bose-Einstein condensate (BEC) is interacting with the quantum superimposed magnetic field of the flux-qubit (FQ). An uniform magnetic field is applied along the quantization axis and it also biases the FQ such that the potential energy of the FQ corresponds to a symmetric double well. The expectation value of the predominant component of the magnetic field, $\langle B_{z}(r)\rangle$, increases with the distance ($z$) for the clockwise flow of the persistent current and decreases with the distance ($z$) for the anti-clockwise flow of the persistent current. After interaction with the magnetic field of the FQ, the BEC can be path entangled.}}
\end{center}
\end{figure}
\end{center}

The quantum SG thought experiment, in addition to the quantization of intrinsic angular momentum of the atom, treats the magnetic field quantum mechanically by incorporating the quantum superposition principle. The quantum superposition of the magnetic field is produced by a flux-qubit (FQ) \cite{legget1, fried, caspar, nori}, which also creates a quantum superimposed magnetic field gradient. The  schematic of a quantum SG experiment is illustrated in Fig.~\ref{fig1}, where the source of neutral atoms is a trapped BEC. If a single atom is un-trapped from the trap then it falls freely under gravity along the positive $x$-direction. Consider the quantum state of the spin degrees of freedom of the atom to be a quantum superposition of the spin projections \emph{i.e.} $\sum_{m_{F}=-F}^{F} c_{m_{F}}(t) |m_{F}\rangle$. The free falling atom comes in the close proximity of the FQ where it interacts, for a time duration $\Delta t$, with the magnetic field produced by the FQ. The mass of the FQ is assumed to be very high and it remains at rest during the interaction.
The atom continues free fall and as it moves away from the FQ, the interaction between the magnetic field and the atom diminishes. The centre of the closed loop of the FQ is considered to be the origin, where the quantization axis is perpendicular to the plane of the FQ loop along the $z$-axis as shown in Fig.~\ref{fig1}. The FQ is a superconducting loop interrupted by a single Josephson-junction, where the net magnetic flux passing through the FQ loop is considered to be the macroscopic quantum observable. The Hamiltonian of the FQ is written as,
$H_{Q}=\frac{p^{2}_{\Phi}}{2 C_{j}}+\frac{(\Phi-\Phi_{a})^{2}}{2L}+E_{j}(1-\cos(2\pi \Phi/\Phi_{o}))$\cite{legget1, revwendin, revshnir, devoret, leggetrev}, Where $\Phi_{o}=h/2 e$ is the magnetic flux quantum, $e$ is electron charge, $\Phi$ is the net magnetic flux passing through the FQ loop, $p_{\Phi}=-i\hbar \partial/\partial \Phi$ is the momentum operator conjugate to $\Phi$, $(\Phi-\Phi_{a})^{2}/2L$ is the magnetic energy stored in the FQ loop of self-inductance $L$, $E_{j}(1-\cos(2\pi \Phi/\Phi_{o}))$ is the potential energy of the Josephson-junction of junction capacitance $C_{j}$, Josephson energy $E_{j}=I_{c}\hbar/2 e$ and $I_{c}$ is the maximum current that can pass through the Josephson-junction without dissipation. The total potential energy of the FQ has two global minima, if the externally applied magnetic flux $\Phi_{a}$ is equal to half of the magnetic flux quantum $(\Phi_{a}=\Phi_{o}/2)$. Therefore, if $\Phi_{a}=\Phi_{o}/2$,  the potential energy profile close to the minima can be considered as a symmetric double-well potential. The potential energy profile becomes asymmetric, if the externally  applied magnetic flux deviates from $\Phi_{o}/2$. The tunneling amplitude between the wells is governed by the barrier height $E_{j}$, which can be controlled through an additional external magnetic field by replacing the single Josephson-junction with a dc-Superconducting Quantum Interference Device (dc-SQUID). Therefore, by allowing the tunneling between the potential wells, the FQ can be prepared in the ground state of the symmetric double-well potential. The ground state of such a symmetric double well potential corresponds to a quantum superposition of the persistent current flowing in the clockwise and in the anti-clockwise direction. Therefore, a quantum superposition of the persistent current flowing in the opposite directions through the FQ loop produces a macroscopic quantum superposition of the magnetic flux. For a further reference, an additional dimension of the FQ has been explored in Ref \cite{fqc}.

The ground-state wave function of the FQ is de-localised over the double-well potential therefore, the persistent current and the corresponding net magnetic flux passing through the loop of the FQ are quantum superimposed. It is the quantum superposition of the magnetic flux which eventually produces a quantum superimposed magnetic field gradient.
Consider an atom is far from the FQ such that the interaction between the atom and the FQ is zero. The FQ is prepared in its ground quantum state $\int C_{g}(\Phi, t)|\Phi\rangle \mathrm{d}\Phi$, where the quantum state is written in the net magnetic flux basis and $C_{g}(\Phi, t)$ is the corresponding ground state wave function. The quantum state of a free falling atom in the centre of mass frame is
$|\psi(\mathbf{r};t)\rangle\sum_{m_{F}=-F}^{F} c_{m_{F}}(t) |m_{F}\rangle$. Therefore, prior to the interaction the total quantum state of the atom and the magnetic flux is a product state
 \emph{i.e.} $|\psi(\mathbf{r};t)\rangle\sum_{m_{F}=-F}^{F} c_{m_{F}}(t) |m_{F}\rangle~\int C_{g}(\Phi, t)|\Phi\rangle \mathrm{d}\Phi$.
During the free fall, the atom comes in the close proximity of the FQ and it interacts, for a time $\triangle t$, with the quantum superimposed magnetic field of the FQ. The interaction vanishes as the free falling atom moves away from the FQ. For each value of the net magnetic-flux, there is a corresponding magnetic field gradient and a corresponding imparted transverse momentum. It is assumed that the net magnetic flux and the magnetic field gradient can be determined simultaneously. Therefore, after the interaction, the total quantum state of the atom and the field becomes
\begin{multline}
\label{ent3}
  |\alpha (\mathbf{r},\mathbf{\Phi},\mathbf{m_{F}};t)\rangle_{T}= \\ \sum_{m_{F}=-F}^{F} \left(\int \int \Psi(r,\Phi, m_{F},t)C_{g}(\Phi,t) |r\rangle |\Phi\rangle \mathrm{d}r\mathrm{d}\Phi \right)\times \\
  c_{mF}(t)|m_{F}\rangle
\end{multline}
In general, the amplitude  $\Psi(r, \Phi, m_{F},t)$ is an inseparable function of  $r$, $\Phi$ and $m_{F}$ therefore, the quantum state given in Eq.~\ref{ent3} is an entangled quantum state. For the quantum state given in Eq.~\ref{ent3} to be treated as an entangled quantum state, the transverse linear momentum uncertainty of the initial wave function $\psi(r,t)$ of the atom should be less than the uncertainty of the transverse linear momentum imparted to the atom \emph{i.e.} the uncertainty of the momentum of wave function $\Psi(r,\Phi,m_{F},t)$ along the transverse direction, which is parallel to the quantization axis. Such an entangled quantum state is a macroscopic hybrid quantum state where, the external degrees of freedom of the atom and the magnetic-flux are continuous variables, while the spin degree of freedom of the atom is discrete. For a given spin projection state $|m_{F}\rangle$, which has a nonzero interaction with the magnetic field, the term in the bracket of Eq.~\ref{ent3} is an entangled quantum state of the atomic path with the magnetic flux.

\section{Bose-Einstein condensate and quantum Stern-Gerlach experiment}
Consider a schematic of a quantum SG experiment as shown in Fig.~\ref{fig1} where a BEC of $N$-atoms is undergoing a free fall and it interacts with the magnetic field of the FQ, which is prepared in its ground state. The BEC is spin polarised, where all atoms of the BEC are prepared in a given spin projection state $|m_{F}\rangle$, which has a nonzero coupling with the magnetic field. The interaction among the atoms of a free falling BEC is assumed to be zero. The quantum state of $N$-atom BEC prior to its interaction with the FQ can be written as $\left(\int \psi(r,t) |r\rangle  \mathrm{d}r\right)^{\otimes N}\equiv \int \psi(r_{1},t) |r_{1}\rangle  \mathrm{d}r_{1}\otimes \int \psi(r_{2},t) |r_{2}\rangle  \mathrm{d}r_{2}.....\otimes \int \psi(r_{n},t) |r_{n}\rangle  \mathrm{d}r_{n}$ where the variables $r_{1}$, $r_{2}$ upto $r_{n}$ are the spatial coordinates of $N$ atoms with the  same wave function $\psi(r,t)$.
Since all the atoms of the BEC are influenced by the superimposed magnetic field therefore, after interaction with the FQ, the total quantum state of the BEC becomes entangled with the quantum state of the FQ. Therefore, the total quantum state can be written as
\begin{equation}
\label{ent4}
  |\beta(\mathbf{r},\mathbf{\Phi};t)\rangle \propto \int \left(\int \Psi(r,\Phi,t) |r\rangle  \mathrm{d}r\right)^{\otimes N} C_{g}(\Phi,t) |\Phi\rangle \mathrm{d}\Phi
\end{equation}
Where, $\left(\int \Psi(r,\Phi,t) |r\rangle  \mathrm{d}r\right)^{\otimes N}\equiv \int \Psi(r_{1},\Phi,t) |r_{1}\rangle  \mathrm{d}r_{1}\otimes \int \Psi(r_{2},\Phi,t) |r_{2}\rangle  \mathrm{d}r_{2}
.....\otimes\int \Psi(r_{n},\Phi,t) |r_{n}\rangle  \mathrm{d}r_{n}$ is the quantum state of noninteracting $N$-atom BEC entangled with the net magnetic flux passing through the loop of the FQ and $C_{g}(\Phi,t)$ is the ground-state  wave function of the FQ. The spin degree of freedom is omitted from Eq.~\ref{ent4} since the BEC is spin polarised. For a fixed value of the net magnetic flux passing through the FQ loop, there is a corresponding imparted transverse linear momentum, which is the same for all atoms of the BEC. It is the imparted transverse linear momentum which is well defined for a given value of the net magnetic flux passing through the FQ loop however, there is a finite uncertainty in the transverse linear momentum due to a finite extension of the wave function of the BEC.
It is important to note that in the case of a semi-classical SG experiment the spin polarised atoms, in a given quantum state $|m_{F}\rangle$, travel along a unique path (with a nonzero uncertainty), while in the case of the quantum SG experiment, even the spin polarised atoms can travel along different distinct paths due to the quantum nature of the magnetic field.

The ground-state of the symmetric double-well potential of the FQ can be written as $(|\Phi\rangle_{L}+|\Phi\rangle_{R})/\sqrt{2}$, where $|\Phi\rangle_{L}$ and $|\Phi\rangle_{R}$ are the quantum states corresponding to the persistent current flowing in clockwise and anti-clockwise directions, respectively. Each potential well of the double-well potential is harmonic around their respective minima, therefore the corresponding ground-state wave function $C_{g}(\Phi,t)$ of the double-well potential can be written as a sum of two Gaussian functions centered at the magnetic flux values $\Phi_{L}$ and $\Phi_{R}$, such that $C_{g}(\Phi)\simeq (\mathbf{e}^{-\frac{(\Phi-\Phi_{L})^2}{2\Delta\Phi^{2}}}+\mathbf{e}^{-\frac{(\Phi-\Phi_{R})^2}{2\Delta\Phi^{2}}}) /\sqrt{2}\pi^{1/4} \Delta\Phi^{1/2}$, where $\Delta\Phi$ is the width of each Gaussian. If the width $\Delta\Phi$ is much less than the distance between the peaks of the wave fuction, \emph{i.e.} $\Delta\Phi\ll \Phi_{R}-\Phi_{L}$, then  $|\Phi\rangle_{R}$ and $|\Phi\rangle_{L}$ can almost be considered orthogonal to each other \emph{i.e.} $_{L}\langle\Phi|\Phi\rangle_{R}\simeq0$.
For a FQ, consisting of a circular super-conducting closed loop, the predominant component of magnetic field and the magnetic field gradient is along the $z$-axis. Consider the expectation value of the predominant component of the magnetic field gradient corresponding to the magnetic flux ground state $|\Phi\rangle_{L}$ of the left potential well and the magnetic flux state $|\Phi\rangle_{R}$ of the right potential well are $\partial \langle B_{z}(r)\rangle_{L}/\partial z$ and $\partial \langle B_{z}(r)\rangle_{R}/\partial z$, respectively.  The magnetic field and its gradient are time dependent with respect to the frame of reference of a free falling BEC. Therefore, after interaction with the magnetic field of the FQ, for a time duration $\Delta t$, the expectation value of the imparted transverse linear momenta (along the $z$-axis) of each atom of the BEC corresponding to $|\Phi\rangle_{L}$ and $|\Phi\rangle_{R}$ are $\langle p_{z}\rangle_{L} = - m_{F}g_{F}\mu_{B} \int^{\Delta t}_{0}\frac{\partial \langle B_{z}(r,t)\rangle_{L}}{\partial z} \mathrm{dt}$ and $\langle p_{z}\rangle_{R} = - m_{F}g_{F}\mu_{B} \int^{\Delta t}_{0} \frac{\partial \langle B_{z}(r,t)\rangle_{R}}{\partial z} \mathrm{dt} $, respectively.
The finite spread $\Delta \Phi$ of the magnetic flux quantum state of each well of the double well of the FQ produces an uncertainty $\Delta P_{z}$ in the imparted transverse linear momentum.
 This uncertainty in the imparted transverse momenta can be neglected if $\Delta\Phi\ll \Phi_{R}-\Phi_{L}$ such that $\Delta P_{z}\ll \Delta p_{z}$, where $\Delta p_{z}$ is the uncertainty of the $z$-component of the linear momentum of the BEC wave function prior to its interaction with the FQ.

\section{Regimes of coupling}
The magnetic field from the FQ also interacts with the environment, which consists of a substrate on which the FQ is fabricated and the measurement devices. Such a coupling with the environment produces decoherence of the FQ quantum state due to its entanglement with the environment.  Suppose an FQ is prepared in its ground state and immediately a BEC enters the region of interaction. To produce a quantum entanglement of the BEC with the magnetic flux only, the interaction of the BEC with the FQ should complete prior to the decoherence of the FQ quantum state \emph{i.e.} the interaction time $\Delta t$ must be considerably less than the decoherence time $t_{d}$. The time of interaction $\Delta t$ is determined by the $x$-component of the velocity of the BEC.
Therefore, three regimes of coupling between BEC and an FQ can be classified as $\mathbf{(1.)}$ The regime of strong coupling, if $|\langle p_{z}\rangle_{R}-\langle p_{z}\rangle_{L}|\gg \Delta p_{z}$ where, $|\langle p_{z}\rangle_{R}-\langle p_{z}\rangle_{L}|=|m_{F}g_{F}\mu_{B}\left(\int^{\Delta t}_{0}\frac{\partial \langle B_{z}(r,t)\rangle_{L}}{\partial z} \mathrm{dt}-\int^{\Delta t}_{0}\frac{\partial \langle B_{z}(r,t)\rangle_{R}}{\partial z} \mathrm{dt}\right)|$. In this case, the BEC is quantum entangled with the quantum state of the FQ \emph{i.e.} the total quantum state given in Eq.~\ref{ent4} is considered to be a macroscopic entangled quantum state. The interaction time $\Delta t$ cannot be increased arbitrarily, in order to increase $|\langle p_{z}\rangle_{R}-\langle p_{z}\rangle_{L}|$, as the interaction time should be less than the decoherence time $t_{d}$. $\mathbf{(2.)}$ The regime of weak coupling, if $|\langle p_{z}\rangle_{R}-\langle p_{z}\rangle_{L}| \sim \Delta p_{z}$. In the case of weak coupling, the wave functions of the BEC with expectation values of the imparted transverse momenta $\langle p_{z}\rangle_{R}$ and $\langle p_{z}\rangle_{L}$ are nonorthogonal \emph{i.e.} the split paths of the BEC are partially overlapping. $\mathbf{(3.)}$ If $|\langle p_{z}\rangle_{R}-\langle p_{z}\rangle_{L}| \ll \Delta p_{z}$, the total quantum state given in Eq.~\ref{ent4} remains a product state.

The ground state of a symmetrically biased FQ is $|\Phi\rangle_{g}=(|\Phi\rangle_{L}+|\Phi\rangle_{R})/\sqrt{2}$. If the uncertainty $\Delta P_{z} \ll \Delta p_{z}$ for $\Delta \Phi\ll \Phi_{R}-\Phi_{L}$, then the quantum state given in Eq.~\ref{ent4} can be approximately written as
\begin{equation}
\label{ent6}
  |\beta(\mathbf{r},\mathbf{\Phi};t)\rangle\approx \frac{1}{\sqrt{2}}\left(|\Psi(\mathbf{r};t)\rangle_{1}|\Phi\rangle_{L}+|\Psi(\mathbf{r};t)\rangle_{2}|\Phi\rangle_{R}\right)
\end{equation}
Where, $|\Psi(\mathbf{r};t)\rangle_{1}$ and $|\Psi(\mathbf{r};t)\rangle_{2}$ are the momentum imparted quantum states of the noninteracting $N$-atom BEC such that

\begin{equation}
\begin{split}
\label{ent7}
& |\Psi(\mathbf{r};t)\rangle_{1}=\left(\int \Psi(r,\Phi_{L},t) |r\rangle  \mathrm{d}r\right)^{\otimes N} \\
  & |\Psi(\mathbf{r};t)\rangle_{2}=\left(\int \Psi(r,\Phi_{R},t) |r\rangle  \mathrm{d}r\right)^{\otimes N}
\end{split}
\end{equation}

\section{Path entanglement of Bose-Einstein condensate}
The interaction between the BEC and the magnetic field of the FQ diminishes as the free falling BEC moves away from the FQ. Consider, immediately after the interaction time $\Delta t$, a Hadamard operation is applied on the FQ quantum state such that $|\Phi\rangle_{L}\mapsto\frac{|\Phi\rangle_{L}+|\Phi\rangle_{R}}{\sqrt{2}}$ and $|\Phi\rangle_{R}\mapsto\frac{|\Phi\rangle_{L}-|\Phi\rangle_{R}}{\sqrt{2}}$. If the time duration to apply a Hadamard operation is $t_{h}$ then $\Delta t +t_{h}$ should be much less than the decoherence time $t_{d}$. Therefore, the quantum state given in Eq.~\ref{ent6}, after applying the Hadamard operation on the FQ quantum state, becomes
\begin{multline}
\label{ent8}
  |\beta(\mathbf{r},\mathbf{\Phi};t)\rangle \approx \frac{1}{\sqrt{2}}\left(\frac{|\Psi(\mathbf{r};t)\rangle_{1}+|\Psi(\mathbf{r};t)\rangle_{2}}{\sqrt{2}}\right)|\Phi\rangle_{L}\\
  + \frac{1}{\sqrt{2}}\left(\frac{|\Psi(\mathbf{r};t)\rangle_{1}-|\Psi(\mathbf{r};t)\rangle_{2}}{\sqrt{2}}\right)|\Phi\rangle_{R}
\end{multline}

After the Hadamard operation, the FQ quantum state is measured in the magnetic flux basis. If the time required to perform this measurement is $t_{m}$ then $\Delta t+t_{h}+t_{m} \ll t_{d}$ \emph{i.e.} the quantum state of the FQ must be measured in a time much less than its decoherence time. Therefore, if the measurement outcome is $|\Phi\rangle_{L}$, then the quantum state of atoms collapses to a path entangled quantum state of a BEC \emph{i.e.} $\frac{|\Psi(\mathbf{r};t)\rangle_{1}+|\Psi(\mathbf{r};t)\rangle_{2}}{\sqrt{2}}$. In a path entangled quantum state, the BEC of $N$-atoms behaves as a single particle whose paths are quantum superimposed. On the other hand, if the measurement outcome is $|\Phi\rangle_{R}$, then the quantum state of atoms collapses to a path entangled BEC of quantum state $\frac{|\Psi(\mathbf{r};t)\rangle_{1}-|\Psi(\mathbf{r};t)\rangle_{2}}{\sqrt{2}}$. If the FQ quantum state is not measured and correlated with the quantum state of the BEC or if the quantum state of the FQ is ignored then the quantum state of the BEC shall be an incoherent mixture of path entangled states with plus and minus signs, which will result in a mixed quantum state, since there is no information available to distinguish them from each other.

After completion of a measurement on the FQ, the BEC is disentangled with the FQ and the path entangled BEC continues a free fall under gravity. Since the atoms are falling in the interaction free region therefore, the path entanglement of BEC persists even for a time much larger than the decoherence time of the FQ. The path entanglement of BEC can be detected by recombining the paths of the entangled BEC and by detecting all the atoms at a given location. As the position of the number detector is displaced an interference pattern can be obtained. Each time exactly the same number of atoms must be prepared in the path entangled state and all of the atoms must be detected at a given position of detector \cite{detect}. However, the resulting interference pattern shall be contracted as compared to the interference pattern of two overlapping BEC as if the de Broglie wavelength of path entangled atoms is reduced such that $\lambda_{path}=\lambda_{BEC}/N$. Where, $\lambda_{BEC}$ is the instantaneous wavelength of an atom from a falling BEC. In the case of path entangled BEC, all of the $N$ noninteracting atoms behave like a single particle whose mass is $N$ times the mass of an individual atom.

\section{Estimations}
To estimate the order of the decoherence time required to produce a strong coupling, consider an FQ of a circular loop cross section, where the inner radius of the loop is 2.0~$\mu$m, the outer radius of the loop is 2.5~$\mu$m and the thickness of the loop is 1.0~$\mu$m. If the loop is fabricated on a nonmagnetic material then the self-inductance of the loop with given dimensions is $L \simeq$ 6.44~pH. An uniform magnetic field, 650$\times$10$^{-4}$~mT, is applied along the $z$-axis to magnetically bias the FQ at a half of the flux quantum, where the flux quantum $\Phi_{o}=$ 2.0678$\times$10$^{-15}$~Tm$^{2}$. Consider minima of the symmetric double-well potential corresponding to the left and the right potential wells are located at 0.25$\Phi_{o}$ and 0.75$\Phi_{o}$, respectively. For these minima values, the difference between the peaks ($\Phi_{R}-\Phi_{L}$) of the ground state wave function is about $\Phi_{o}/2$. Assuming, $\Delta\Phi\ll \Phi_{R}-\Phi_{L}$, therefore, for the quantum state $|\Phi\rangle_{L}$, corresponding to the left potential well, the expectation value of the persistent current is -80.3~$\mu$A, which produces a magnetic field in the opposite direction to the applied bias magnetic field. Similarly, for the quantum state $|\Phi\rangle_{R}$, corresponding to the right potential well, the expectation value of the persistent current is +80.3~$\mu$A and it produces a magnetic field in the direction of the applied bias magnetic field. The Josephson junction critical current should be higher than the calculated expectation value of the persistent current. Corresponding to the ground state of the left potential well, the maximum of the expectation value of the magnetic field gradient is $\frac{\partial \langle B_{z}(r,t)\rangle_{L}}{\partial z} \sim$~81.8$\times$10$^{-1}$~Tm$^{-1}$ at a distance  $\sim$~1.25~$\mu$m from the centre of the loop on the $z$-axis. Corresponding to the ground state of the right potential well, the direction of persistent current is reversed therefore, the minimum of the expectation value of the magnetic field gradient is $\frac{\partial \langle B_{z}(r,t)\rangle_{R}}{\partial z} \sim$~-81.8$\times$10$^{-1}$~Tm$^{-1}$ at the same distance $\sim$~1.25~$\mu$m from the centre of the loop on the $z$-axis. Around this point, the magnetic field gradient remains almost constant up to a distance of about 2.0~$\mu$m in a plane parallel to the plane of the loop and up to a distance of about 1.0~$\mu$m along the $z$-axis. Therefore, for an anisotropic BEC just before it enters the magnetic field region of the FQ the Gaussian wave function of the falling BEC has widths along the directions parallel to the $z$-axis and the $y$-axis to be 1.0~$\mu$m.
The free falling  BEC can be guided through an atom waveguide up to the interaction region in order to maintain its required extension in the $y$-$z$ plane prior to its interaction with the FQ. The waveguide potential can be turned off immediately when the BEC enters in the interaction region.
The direction of the velocity of the BEC during the free fall is parallel to the $x$-axis. The width of the BEC along a direction parallel to the $x$-axis is considered to be 5.0~$\mu$m. The position of the FQ is adjusted such that during the free fall the BEC passes through the region of maximum magnetic field gradient, which is located at a distance $\sim$~1.25~$\mu$m from the centre of the loop. The momentum uncertainty of the BEC, prior to its interaction with the FQ, along the quantization axis ($z$-axis) is $\Delta p_{z}$=$\hbar/\Delta z$, where $\Delta z$ is the width of the wave function of the BEC along a direction parallel to the $z$-axis. Therefore, in order to achieve the strong coupling regime \emph{i.e.} $|\langle p_{z}\rangle_{R}-\langle p_{z}\rangle_{L}|\gg \Delta p_{z}$, the calculated time of interaction $\Delta t$ between the BEC and the FQ should be much greater than $\sim$~2.0~$\mu$sec. Where, the time limit, $\sim$~2.0~$\mu$sec, is calculated for the weak coupling regime for a finite extension of a BEC along the $x$-direction. A BEC of rubidium atoms ($^{87}$Rb) in a quantum state $F$ = $2$ and $m_{F}$ = $+2$ ($g_{F}$ = $0.5$) is considered in the calculations.
 The $x$-component of the velocity of the free falling BEC is chosen to be such that the condensate remains in the high magnetic field gradient region for a time $\Delta t$, which should be much less than the decoherence time.
The decoherence time of an FQ of the order of micro-seconds has been observed \cite{nori, you}. However, the FQ with dimensions described in this paper is considered to have a low self-inductance, which results in a high expectation value of the persistent current $\sim$~80.3~$\mu$A. The Josephson-junction of the FQ should be able to pass through it more than the calculated value of the persistent current without any dissipation.  Furthermore, to realize a path entangled BEC, the decoherence time of the FQ should be such that $t_{d}\gg$2.0~$\mu$sec$+t_{h}+t_{m}$.

\section{Back-action of atoms on Flux-Qubit}
The back action of atoms on the FQ can be classified in two categories, where $\mathbf{(1.)}$ If the mass of the FQ is finite then it should be displaced to conserve the total linear momentum during the interaction. In this paper, the mass of the FQ is considered to be very high as compared to the total mass of atoms in the BEC. Therefore, FQ remains stationary. $\mathbf{(2.)}$ The magnetic moment of atoms is nonzero therefore, each atom produces its own intrinsic magnetic field. This intrinsic magnetic field can change the net magnetic flux linked to the FQ if the atoms are situated very close to the FQ. This effect is  predominant in the case of a spin polarised BEC of a large number of atoms which can produce a considerable magnetic field to deviate the bias of the FQ from one-half of the flux quantum. The variation in the bias of the FQ due to the magnetic field of atoms can modify the potential energy and hence the quantum state of the FQ during the interaction, which is otherwise considered to be the same throughout the interaction. To estimate the back-action, consider an atom as an ideal magnetic dipole situated on the $z$-axis at a distance $z_{o}$ with its spin pointing along the $z$-axis.  The center of the circular FQ loop of radius $R$ is located at the origin and the loop is situated in the $x$-$y$ plane. In the case of an atom spin polarised along the $z$-axis the $z$-component of the magnetic field due to its magnetic moment is $(\mu_{o}m_{F}g_{F}\mu_{B}/4\pi)(3z_{o}^{2}/(x^{2}+y^{2}+z_{o}^{2})^{5/2}-1/(x^{2}+y^{2}+z_{o}^{2})^{3/2})$ and the corresponding magnetic flux linked to the circular loop is $\Phi_{atom}=(\mu_{o}m_{F}g_{F}\mu_{B}/2)(R^{2}/(R^{2}+z_{o}^{2})^{3/2})$ where, $\mu_{o}$ is the magnetic permeability of the free space. For a spin polarised BEC of $N$ atoms the magnetic flux linked to the FQ loop due to the magnetic field of $N$ atoms is $N \Phi_{atom}$ where, all the atoms are assumed to be located at the same point. The back action of $N$ atoms on the FQ is negligible if the magnetic flux linked to the FQ loop due to atoms is much lower than the flux quantum $\Phi_{o}$ such that $N\Phi_{atom}/\Phi_{o}\ll 1$. In this case the atoms will deviate the bias magnetic flux of the FQ to a negligible extent. Consider a FQ loop of zero thickness of radius $R = 2.25~\mu m$. A free falling spin polarised BEC passes through the point $z_{o}=1.25~\mu$m where the field gradient is close to the maximum as considered in the previous section. A typical number of atoms in a BEC is $\sim 10^{5}$ and if all atoms are spin polarised and are momentarily located at $z_{o}$ during the free fall, the maximum value of the parameter $N\Phi_{atom}/\Phi_{o}\simeq 8.4\times 10^{-5}$. Therefore, the FQ remains in its initial ground state during the interaction with BEC and its initial ground state becomes entangled with the atoms of BEC. The number of atoms should be more than $\sim 10^{8}$ in order to produce a considerable effect of back-action. Therefore, for a typical number of atoms $\sim 10^{5}$-$10^{6}$ the back-action of BEC on the FQ is negligible and the decoherence rate of the BEC-FQ quantum system is determined by its interaction with the environment.

\section{Conclusion}
In this paper, a quantum SG thought experiment and its fundamental significance has been presented. In addition to the intrinsic angular momentum of an atom the magnetic field is also treated quantum mechanically by incorporating the quantum superposition principle. As a consequence, the path of atoms can split into more than one distinct paths even for the case of spin polarised atoms along the quantization axis.
In contrast to the semi-classical SG experiment, the quantum SG experiment can produce a macroscopic quantum entanglement of the path of a BEC with the quantum state of the magnetic flux. In addition, the quantum SG experiment can produce a path entanglement of BEC where, the BEC can occupy physically distinct locations. The path entanglement of the BEC can persist for a time much larger than  the decoherence time of the FQ. Three different regimes of coupling between the FQ and  BEC are also discussed. A measure of back-action of a BEC on the FQ is introduced in the last section and for atom number $N \sim 10^{5}$ the back-action is negligible.

\acknowledgments{Author is thankful to Prof. Arvind for suggestions about the paper.}

\bibliography{aps}
\end{document}